\documentclass[aps,nofootinbib,notitlepage,showpacs,longbibliography]{revtex4-1}

\usepackage[CJKbookmarks, pdftex, bookmarksnumbered, bookmarksopen, colorlinks, citecolor=blue, linkcolor=blue]{hyperref}

\usepackage{amstext,amsmath,amssymb,amsfonts,bbm}
\usepackage[latin1]{inputenc}
\usepackage{graphicx}
\usepackage{epstopdf}
\usepackage{amsthm}
\usepackage{tocvsec2}
\usepackage{enumerate}
\usepackage{subfigure}
\usepackage{color}

\usepackage{multirow} \usepackage{rotating}

\def\beq{\begin{equation}}
\def\be{\begin{equation}}
\def\ee{\end{equation}}
\def\bes{\begin{eqnarray}}
\def\ees{\end{eqnarray}}

\begin{document}

\title{How many quanta are there in a quantum spacetime?}

\author{Seramika Ariwahjoedi$^{1,3}$, Jusak Sali Kosasih$^{3}$, Carlo Rovelli$^{1,2}$, Freddy P. Zen$^{3}$\vspace{1mm}}

\affiliation{$^{1}$Aix Marseille Universit\'e, CNRS, CPT, UMR 7332, 13288 Marseille, France.\\
$^{2}$Universit\'e de Toulon, CNRS, CPT, UMR 7332, 83957 La Garde, France.\\$^{3}$Institut Teknologi Bandung, Bandung 40132, West Java, Indonesia.}

\begin{abstract} 

\noindent
Following earlier insights by Livine and Terno, we develop a technique for describing quantum states of the gravitational field in terms of \emph{coarse grained} spin networks.  We show that the number of nodes and links and the values of the spin depend on the observables chosen for the description of the state. Hence the question in the title of this paper is ill posed, unless further information about what is been measured is given.
\end{abstract}

\maketitle

\section{Introduction}

The electromagnetic field can be viewed as formed by individual photons. This is a consequence of quantum theory.  Similarly, quantum theory is likely to imply a granularity of the gravitational field, and therefore a granularity of space \cite{Rovelli1994a,Ashtekar:1996eg}.  How many  quanta form a macroscopic region of space?   This question has implications for the quantum physics of black holes  \cite{Frodden:2011zz}, scattering calculations in non perturbative quantum gravity \cite{Magliaro2011} and quantum cosmology \cite{Rovelli2008d,Bianchi:2010zs}. It is related to the question of the number of nodes representing a macroscopic geometry in a spin network state in loop gravity \cite{Sahlmann:2001nv}.  In this context, it takes the following form: what is the relation between a state with many nodes and small spins, and a state with few nodes but large spins?  

To gain some insight into this problem, we first review two elementary cases: the number of photons in an electromagnetic field and the number of quanta on coupled oscillators. The two examples throw some light on the problem and reveal the ambiguity of the notion of ``number of quanta". This is done in Section \ref{II}. Then, in Section \ref{III} we define some tools that allow us to talk about coarse-grained observables in quantum gravity. The detailed technique of the proposed coarse-graining procedure is described in detail in Section \ref{IV}, in terms of spin network states. This is inspired by work by Livine and Terno \cite{Livine2008,Livine:2006xk}. Section \ref{V} contains a discussion of statistical properties of the coarse-grained states. In Section \ref{VI} we discuss the geometrical interpretation of the proposed tools. We address the main question in Section \ref{VII}.

\section{How many quanta in a field?} \label{II}

Consider a free scalar field in a finite box, in a classical configuration
$\phi(x,t)$. Can we associate a number of quanta to its state? The
answer is yes, because the standard quantum-field-theoretical number operator,
which sums the number the quanta on each mode, has a well defined classical limit.
The number operator is 
\begin{equation}
N=\sum_{n}N_{n}=\sum_{n}\ a_{n}^{\dagger}a_{n}\label{eq:2-1-1}
\end{equation}
where $a_n$ and $a^\dagger_n $ are the annihilation and creation
operators for the mode $n$ of the field and the sum is over the modes,
namely the Fourier components, of the field. Since the energy can
be expressed as a sum over modes as 
\begin{equation}
E=\sum_{n}E_{n}=\sum_{n}\hbar\omega_{n}\ a_{n}^{\dagger}a_{n}\label{eq:2-2}
\end{equation}
where $\omega_n$ is the angular frequency and $E_n$ its energy of
the mode $n$, it follows that the number of particles is given by
\begin{equation}
N=\frac{1}{\hbar}\ \sum_{n}\frac{E_{n}}{\omega_{n}}\label{eq:2-3}
\end{equation}
which is a well defined classical expression that can be directly
obtained from $\phi(x,t)$ by computing the energy in each mode. Therefore 
each classical configuration defines a total particle-number $N$ and a 
distribution of these particles over the modes 
\begin{equation}
N_{n}=\frac{E_{n}}{\hbar\omega_{n}}.\label{eq:2-4}
\end{equation}
An antenna tuned on the frequency of the mode $n$ absorbs and
emits quanta of the mode $n$ in discrete steps, because of quantum theory. 
Therefore an antenna
interacts with specific modes, and the quanta of these modes express
the quantum discreteness that shows up in this interaction. Notice
that this remains true also if there are interactions. In this case,
the particle number may not be conserved in time, because of the dynamics, but remains
nevertheless a well defined quantity at each time. 

Thus we can compute a ``number of quanta" 
associated to a classical configuration of a field. This number  
is also the expectation value of the number operator in the 
coherent state associated to the given classical field configuration.

This conclusion, nevertheless, must be interpreted with caution, as the following
example shows.  Consider two coupled oscillators $q_1$ and $q_2$ with Hamiltonian
$H=H_1+H_2+V$, where $H_1$ and $H_2$ are free oscillator Hamiltonians
for the two degrees of freedom and $V$ is a coupling. Any state of
this system can be expanded on the basis $|n_1,n_2\rangle_{12}$ that
diagonalizes $H_1$ and $H_2$ and we can define $N=n_1+n_2$ as the
total number of quanta in the state. A detector that measures the
amplitude of the oscillations of $q_1$ can measure the number $n_1$. But we can also expand the variables $q_1$ and $q_2$ into the
two normal modes $q_+$ and $q_-$ which diagonalize the Hamiltonian.
Expanding the state in the basis $|n_+,n_-\rangle_{\pm}$ that diagonalizes
the energy of the two modes, we obtain a \emph{different} definition
$\tilde N=n_++n_-$ of the total number of quanta. Thus the ``number of quanta" depends on \emph{which kind of quanta} one
is considering. A ``one particle state" in the sense
$|1,0\rangle_{12}$ or $|0,1\rangle_{12}$ is \emph{not} a linear combination of ``one particle state" in the sense $|0,1\rangle_{\pm}$ and $|1,0\rangle_{\pm}$, as an elementary calculation may confirm. 

This shows that the ``number of quanta" is a slippery
notion, because it depends on what exactly one is asking. In physical
terms, it depends on how we interact with the system. If we interact
with one of the two oscillators we absorb and emit $n_{1,2}$ quanta;
if we have a device coupled to the modes, we emit and absorb $n_\pm$
quanta. Notice that the field-theory particles are the analog to the
$n_\pm$ quanta, not the $n_{1,2}$ quanta. In this sense they are
non local \cite{Colosi:2004vw}. 

The conclusion drawn from these elementary examples is therefore double: we
can associate a notion of ``number of quanta" to a classical
configuration, but only after we have specified that we are interested
in counting quanta of a specified variable or set of variables. 

Quantum discreteness is not the existence of elementary ``bricks" of nature. It is the appearance  of discreteness in the way a system interacts.  An interaction depends on a variable of the system and this variables my have discrete spectrum. If a different variables interact, different kinds of discreteness (classically incompatible with one another) show up.  With these considerations in mind, let's study the number of quanta in quantum gravity.

\section{Subset graphs} \label{III}

As first observed by Lewandowsky \cite{JerzyGraph}, the state space
of loop quantum gravity contains subspaces ${\cal H}_\gamma$  
associated to abstract graphs $\gamma$. A graph $\gamma$ is
defined by a finite set $\cal N$ of $|{\cal N}|$ elements $n$ called
\textit{nodes} and a set $\cal L$ of $|{\cal L}|$ oriented couples
$l=(n,n')$ called \textit{links}. (For convenience of notation,
we consider also a link with reversed orientation $l^{-1}=(n',n)$
for every link $l=(n,n')$.) ${\cal H}_\gamma$ is a Hilbert space
isomorphic to $L_2[SU(2)^{|{\cal L}|}]\ni\psi(U_{nn'}), U_{nn'}\in SU(2)$.
There is an action of the local gauge group of the theory on this
Hilbert space, given by $\psi(U_{nn'})\to\psi(\lambda_n U_{nn'} \lambda^{-1}_{n'})$
for $\lambda_n\in SU(2)$ \cite{Rovelli2011c}. The states invariant
under this action form the gauge-invariant (proper) subspace ${\cal K}_\gamma$
and we call $\pi_\gamma$ the orthogonal projection from ${\cal H}_\gamma$
to ${\cal K}_\gamma$.

In the following we work also with statistical states.
These are described by positive operators $\rho$ on ${\cal H}_\gamma$,
such that $\textrm{tr}[\rho]=1$. A pure state $|\psi\rangle\in{\cal H}_\gamma$
determines the density matrix $\rho_\psi=|\psi\rangle\langle\psi|$.
A generic state can be written in the form 
\begin{equation}
\rho=\sum_{n}p_{n}|\psi_{n}\rangle\langle\psi_{n}|\label{eq:3-1}
\end{equation}
where $p_n$ is a probability distribution (that is: $0\le p_n\le 1$ and $\sum_n p_n=1$) over a basis $|\psi_n\rangle$.
A state can also be seen as a positive functional on the observable
algebra, given by the expectation value 
\begin{equation}
\rho(A)\equiv\textrm{tr}[\rho A]=\sum_{n}p_{n}\langle\psi_{n}|A|\psi_{n}\rangle.\label{eq:3-2}
\end{equation}

In the loop gravity literature, the study of quantum geometry associated
to the states in ${\cal H}_\gamma$ is well developed \cite{Sahlmann:2001nv,Bianchi:2010gc,Freidel:2010bw}.
The operators defined on ${\cal H}_\gamma$ can be interpreted as the description of the
geometry of $|{\cal N}|$ quantum polyhedra connected to one another
when there is a link between the corresponding nodes. The left invariant
vector field $\vec J_{nn'}$ that acts on the $U_{nn'}$ variable
is interpreted as the normal to the corresponding face of the polyhedron,
normalized to the area of the face. (The right invariant vector field
$\vec J_{n'n}=U_{nn'}\vec J_{nn'}U^{-1}_{nn'}$ acts on the inverse
link and represents the same face measured from the frame of the other polyhedron.) 
We call ${\cal J}_\gamma$ the algebra generated by these operators, together with the (diagonal) operators $U_l$ defined by the group elements themselves.
On a gauge invariant state $\left|\psi_{\textrm{inv}}\right\rangle \in{\cal K}_{\gamma}$
\begin{equation}
\vec{C}_{n}\left|\psi_{\textrm{inv}}\right\rangle \equiv\sum_{n'}\vec{J}_{nn'}\left|\psi_{\textrm{inv}}\right\rangle =0,\label{eq:3-3}
\end{equation}
 where the sum is over the links (and inverse links) that start at
$n$. This equation defines ${\cal K}_\gamma$. The Minkowski
theorem states that this equation is sufficient for the consistency
of the geometric interpretation of each polyhedron \cite{Bianchi:2010gc}.
The area of each face of the polyhedron is $A_{nn'}=8\pi\gamma G|\vec{J}_{nn'}|$,
and the volume associated to each node is $v(J_{nn'})$ where $v$
is (a suitable ordering) of the function giving the classical volume
of the polyhedron. The expression for the volume is well defined for
the states satisfying (\ref{eq:3-3}); it is convenient to extend
it to the whole of ${\cal H}_\gamma$ by sandwiching it between two
projectors 
\begin{equation}
V_{n}=\pi_{n}\ v(J_{n,n'})\ \pi_{n}\label{eq:3-4}
\end{equation}
where 
\begin{equation}
\pi_{n}=\int_{SU(2)}d\lambda_{n}\ \lambda_{n}\label{eq:3-5}
\end{equation}
is the projector on ${\cal K}_\gamma$.

Irrespective of the geometrical interpretation, we can diagonalize
all $A_l$ and $V_n$ on ${\cal H}_\gamma$, because they commute.
They form a complete commuting set of operators on ${\cal K}_\gamma$
and therefore they define a basis on this space, labelled by their quantum
numbers $j_l$ and $v_n$. This basis, denoted by 
\begin{equation}
\left|j_{l},v_{n}\right\rangle _{\gamma}\label{eq:3-6}
\end{equation}
is the spin network basis.

In the loop literature, a relation between state spaces on different
graphs has often been considered, focusing in particular on the case
where a graph $\Gamma$ is subgraph of the graph $\gamma$ and there
is a natural map from ${\cal H}_\Gamma$ to the subset of ${\cal H}_\gamma$
formed by states where $j_l=0$ if $\l\in\Gamma$ but $\l\notin\gamma$.
Here we consider, instead, a different relation between graphs, defined as follows. 

\begin{figure}
\centerline{\includegraphics[height=6cm]{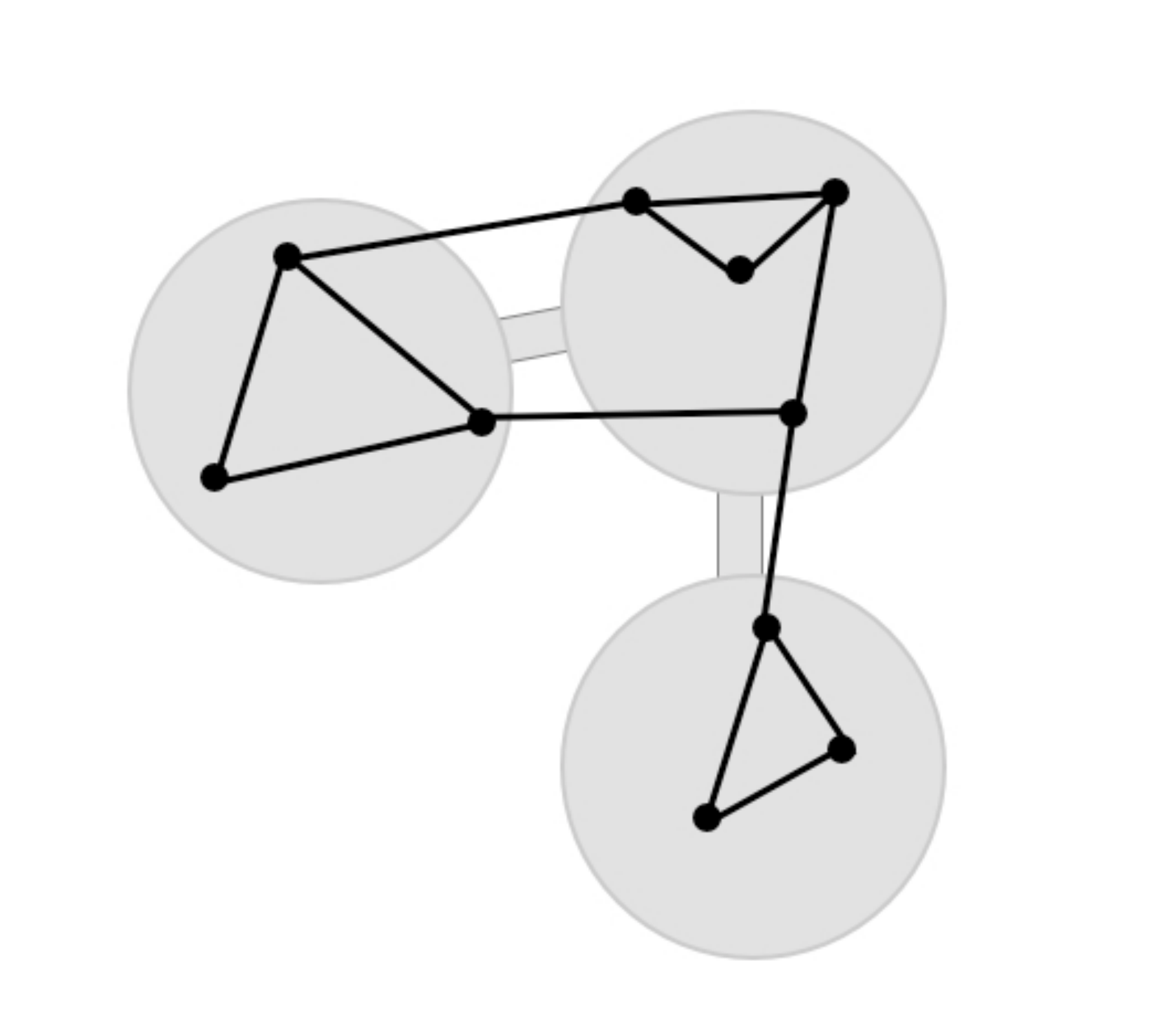}}
\caption{A graph $\Gamma$ in black and a subset graph $\gamma$ in grey.}
\end{figure}

Given a graph $\gamma$, let us define a ``subset graph" $\Gamma$ (not to be confused with a subgraph) as follows. Consider
a partition of ${\cal N}$ into subsets $N=\{n, n',n'', ...\}$, called
``big nodes", such that each $N$ is a connected component
of $\gamma$, namely it is a set of nodes connected among themselves by sequences of links entirely formed by nodes in $N$. See Figure 1. Consider two such big nodes
$N$ and $N'$. Say that they are ``connected" if there
is at least one link of $\gamma$ that links a node in $N$ with a
node in $N'$, and in this case, say that there is a ``big link" $L=(N,N')$ connecting the two. The 
set of the big nodes and the big links defines a graph, which we call ``subset graph" $\Gamma$ 
of $\gamma$. For a small link $l=(n,n')$, we say that $l\in L=(N,N')$ if $n\in N$ and $n'\in N'$.
It is convenient for technical reasons to chose one link $l\in L$ for each $L$ and call it the representative of $L$ in $\gamma$. We assume this choice is part of the definition of the subset graph. 

Now let us construct an algebra of operators in ${\cal H}_\Gamma$,
determined by the subset graph $\Gamma$. For each big link $L$,
let 
\begin{equation}
\vec{J}_{L}:=\sum_{l\in L}\vec{J}_{l}\label{eq:3-7}
\end{equation}
 where the sum is over all links $l=(n,n')$ such that $n\in N$ and
$n'\in N'$. Similarly, for each link $L$ let 
\begin{equation}
U_{L}:=U_{l} \label{eq:3-7-2}
\end{equation}
where $l$ is the representative of the link $L$.
A straightforward calculation shows that the algebra
of these operators, written as ${\cal J}_\Gamma$, is 
\begin{eqnarray} 
[{J}_{L}^{i},{J}_{L'}^{j}]&=&\delta_{LL'}\epsilon^{ij}{}_{k}{J}_{L}^{k},\label{eq:3-8} 
\\     {[} {J}_L^i,, U_{L'}  {]} &= &  \delta_{LL'}\tau^i{U}_{L},\label{eq:3-2}
\\   {[}U_{L},U_{L'} {]}&=&0. \label{eq:3-8-3}   
\end{eqnarray}
This is precisely the algebra ${\cal J}_\Gamma$ of the
operators defined on the Hilbert space ${\cal H}_\Gamma$. Therefore
every state $\rho_\gamma$ on ${\cal H}_\gamma$ determines immediately
a state $\rho_\Gamma=\rho_\gamma|_{{}_{{\cal J}_\Gamma}}$ on ${\cal H}_\Gamma$
by simply restricting it to the algebra of operators ${\cal J}_\Gamma$.  Of course in general $\rho_\Gamma$ is not going to be pure even if $\rho_\gamma$ is. 

We call ${\cal K}_\Gamma$ the proper linear subspace of ${\cal H}_\Gamma$
defined by 
\begin{equation}
\vec{C}_{N}\left|\Psi_{\textrm{inv}}\right\rangle \equiv\sum_{N'}\vec{J}_{NN'}\left|\Psi_{\textrm{inv}}\right\rangle =0.\label{eq:3-9}
\end{equation}
and $\pi_N$ the orthogonal projection on the kernel of $\vec C_N$.
A state on ${\cal H}_\gamma$ can be restricted and projected to a
state in ${\cal K}_\Gamma$. Notice that in general gauge transformations act differently on the graph $\gamma$ and graph $\Gamma$. 

On ${\cal K}_\Gamma$, we define the ``area of the big link" by 
\be
{A}_{L}=8\pi\gamma \hbar G|\vec{J}_{L}|
\ee 
and the ``volume of the big
node" by 
\be
{V}_{N}=\pi_N \ v(\vec J_{NN'})\ \pi_N,
\ee
 where we
recall that $v$ is the expression for the classical volume of a polyhedron.
The operators $\vec{A}_{L}$ and $\vec{V}_{N}$ commute, so they can be diagonalized
together.
%
The quantum numbers
of the big areas are half integers $J_L$ and let the quantum numbers
of the volume be $V_N$. Therefore there is a basis 
\begin{equation}
|J_{L},V_{N},\alpha\rangle\label{eq:3-10}
\end{equation}
where $\alpha$ indicates any other quantum number needed to remove degeneracy. 
Given a state $\left|\psi\right\rangle \in{\cal H}_{\gamma}$, we can construct the corresponding density matrix in ${\cal H}_{\Gamma}$, by 
\begin{equation}
\rho_\Gamma=\textrm{tr}_{\alpha}|\psi\rangle\langle\psi|.\label{eq:3-11}
\end{equation}
and then project it on ${\cal K}_{\Gamma}$ to get a quantum statistical state of the geometry associated to $\Gamma$. 
\begin{equation}
\rho_{\psi}=\pi_\Gamma\ \textrm{tr}_{\alpha}|\psi\rangle\langle\psi|\  \pi_\Gamma .\label{eq:3-11}
\end{equation}
Thus, for any pure state on the fine grained graph, we get a density matrix on the coarse grained graph. 
This construction defines a natural coarse-graining in the space of the spin network states. In the next section we construct coarse-grained states in detail.

\section{Coarse-graining spin networks} \label{IV}

Coarse-graining is a procedure to describe physicals system with 
a smaller number of variables, capturing useful information on the 
system under  a lower resolution.  Coarse-graining is ubiquitous in physics.  
When we describe the motion of a pendulum, or a stone, for instance, we 
are describing the physics of the center of mass, which coarse-grains the 
variables of the individual atoms.
In turn, the physics of an atom is a coarse-grained description where
the individual positions of its quarks are neglected. In field theory,
we work with smeared, namely coarse grained 
observables.  Coarse-grained observables are quantized and can
be discrete as a consequence of quantum theory: the original 
Stern-Gerlach experiment, which clarified the discrete nature of 
angular momentum, for instance, measured the spin of silver atoms: 
namely a coarse grained quantity, not the individual spins of the 
atoms's components. 

Here we describe concretely the possibility of coarse-graining spin network
states. 

\subsubsection{Coarse graining nodes}

\begin{figure}[t]
\centerline{\includegraphics[height=3cm]{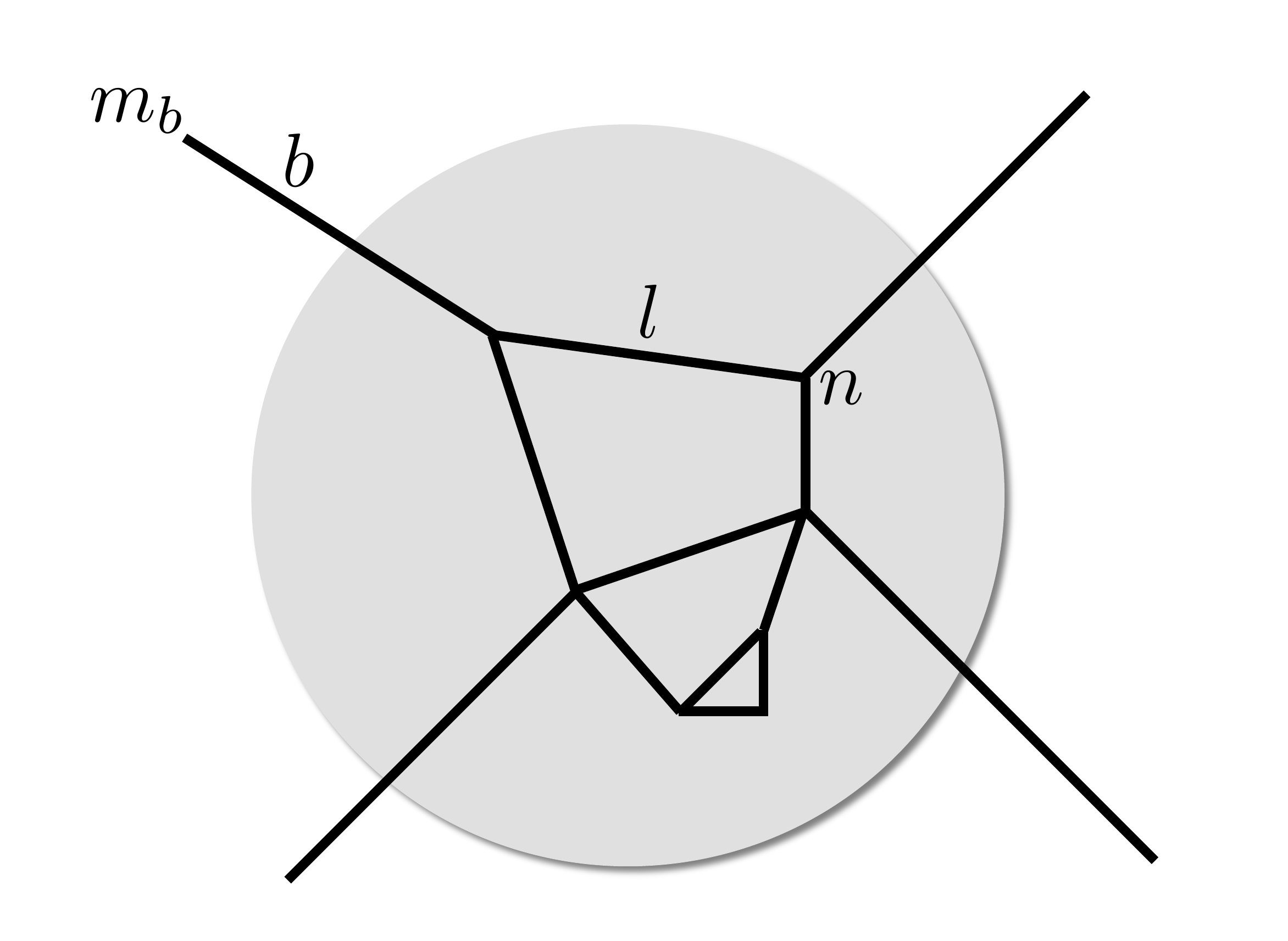}}
\caption{Coarse graining a node.}
\label{nodi}
\end{figure}

We first study the space ${\cal H}_\gamma$  of the non-gauge-invariant states.  This is the space of the square integrable functions on $\left|\mathcal{L}\right|$ copies of $SU(2)$, where $\left|\mathcal{L}\right|$ is the number of links of $\gamma$. The Peter-Weyl theorem states that the Wigner matrices $D^j_{mm'}(U)$ form an orthogonal basis of $L_2[SU(2)]$.  We call the elements of this basis $|j,m,m'\rangle$. That is, we write
\be
D^j_{mm'}(U)=\langle U |j,m,m'\rangle.
\ee
In abstract form, the Hilbert space associated to a single link has the structure 
\be
L_2[SU(2)]=\oplus_{j}(\mathcal{H}_{j}\otimes\mathcal{H}_{j})
\ee
where $\mathcal{H}_{j}$ is the space of the spin-$j$ irreducible representation of $SU(2)$.  It follows that a basis in  ${\cal H}_\gamma$ is given by the quantum numbers $|j_l,m_l,m'_l\rangle$, with $l=1,..., \left|\mathcal{L}\right|$, and $m$ and $m'$ are magnetic quantum numbers in the spin-$j$ representation. Each magnetic numbers $m_l$ (or $m'_l$) of a link $l$ transforms under the  gauge transformations associated to one of the nodes at one end of the link.  Therefore it is naturally associated to one end of the link, namely to a leg $l$ of a node $n$. We can therefore group the magnetic quantum numbers by nodes, and write the basis in the form $|j_l,m_{nl}\rangle$, where $l$ labels the legs of the node $n$.  
\be
|\psi\rangle=\sum_{j_{l},m_{nl}}c_{j_{l},m_{nl}} |{j_{l},m_{nl}}\rangle.
\ee
We consider the general case of an open graph $\gamma$ with external legs ending on an open end (that is, a single-valent node).  Let us distinguish the nodes of the graph into boundary (single-valent) ones, which we denote $b$ and internal ones, which we denote $n$. Similarly, we distinguish the links into the external ones (adjacent to an external node $b$) which we also denote $b$, and the internal ones which we denote $l$. Then
\be
|\psi\rangle=\sum_{j_{b},m_{b},j_{ l},m_{ nl}}c_{j_{b},m_{b},j_{l},m_{nl}} |_{j_{b},m_{b},j_{ l},m_{nl}}\rangle,
\label{purestate}
\ee
where $j_{b}\neq j_{l}$ and $m_{b}\neq m_{nl}$.
Say now that we coarse-grain the entire graph $\gamma$ into a graph $\Gamma$ formed by a single node $N$ with legs $b$. See Figure \ref{nodi}.
  The quantum numbers labelling a basis of the second are simply 
\be
|j_{b},m_{b}\rangle.
\ee
And therefore we can identify the quantum numbers $(j_{l},m_{nl})$ as those corresponding to the $\alpha$'s at the end of the previous section. By tracing over these quantum numbers, the pure state \eqref{purestate} gives
\be
   \rho=\textrm{tr}_{j_{l},m_{nl}} |\psi\rangle\langle\psi|.
\ee
This has matrix elements 
\be
\langle j_{b},m_{b}|\rho|j'_{b},m'_{b}\rangle=\sum_{j_{ l},m_{ nl}}c_{j_{b},m_{b},j_{l},m_{nl}}\overline{c}_{j'_{b},m'_b,j_{l},m_{nl}}.
\ee
Now suppose the state $\psi$ was invariant under gauge transformations on the internal nodes. Then it must be a linear combination of the gauge invariant states $|j_l,m_l,v_n\rangle$ where $v_n$ labels a basis of intertwiners $v_n^{m_{nl}}$ in ${\cal H}_n$. That is:
\be
c_{j_{b},m_{b},j_{l},m_{nl}}=\sum_{v_n} c_{j_{b},m_{b},j_{l},v_n}v_n^{m_{nl}}. 
\ee
Then, using the orthogonality of the intertwiners  
\be
\langle j_{b},m_{b}|\rho|j'_{b},m'_{b}\rangle=\sum_{j_{l},v_n}c_{j_{b},m_{b},j_{l},v_n}\overline{c}_{j'_{b},m'_b,j_{l},v_n}.
\ee
The sum over $v_n$ is always over a finite number of terms, because the ${\cal H}_n$'s have finite dimensions. The sum over the internal links $j_l$ is over a finite or infinite number of terms according to whether the graph $\gamma$ contains loops or not, because if it doesn't the Mandelstam identities make the range of the $j_l$ finite.  The matrix $\rho$ can be projected on the gauge invariant subspace of ${\cal H}_\Gamma$ by contracting it with a basis of on this space, formed by intertwines $v^{m_b}$ on the single node. This gives  
\be
\langle j_{b},v|\rho|j'_{b},v'\rangle=\sum_{j_{l},v_n,m_b,m'_b}c_{j_{b},m_{b},j_{l},v_n}\overline{c}_{j'_{b},m'_b,j_{l},v_n}v^{m_b}v'{}^{m'_b}.
\ee
In particular, the basis state $| j_{l},v_n,j_b,m_b \rangle $ is coarse grained to the density matrix
\be
\rho_{(j_{l},v_n,j_b,m_b)}=\sum_{v\,v'}v^{m_b}v'{}^{m_b}|j_{b},v'\rangle   \langle j_{b},v|.
\ee

If we write explicitly the state on the small graph in the group element basis $\psi(U_l,U_b)$, the coarse grained density matrix turns out to be given explicitly by 
\be
\langle j_{b},v|\rho|j_{b'},v'\rangle=
v^{m_b}v'{}^{m_{b'}}
\int dU_b dU_{b'}\  dU_l \ 
D(U_b)^{j_b}_{m_b n_l}
D(U_{b'})^{{j}_{b'}}_{m_{b'}{n}_{l}}
\overline{\psi(U_l,U_b)}\ \psi(U_{l},U_{b'}). 
\ee
This gives explicitly the gauge invariant density matrix of a large node for any gauge invariant state of the small graph.

\subsubsection{Coarse graining links}

\begin{figure}[t]
\centerline{\includegraphics[height=2cm]{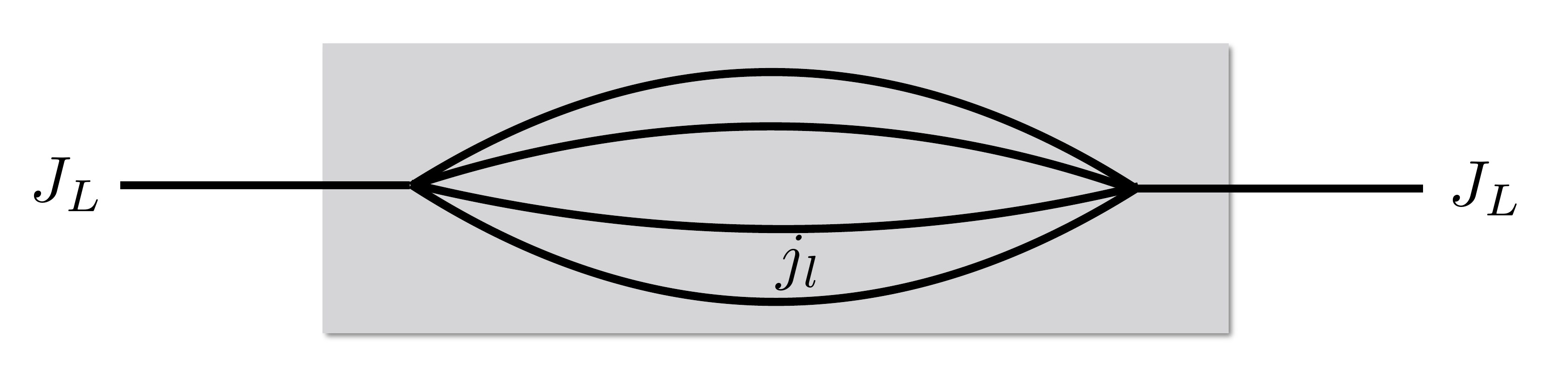}}
\caption{Coarse graining a link.}
\label{links}
\end{figure}

Let us now consider a set of small links $l$ that are contained in a single large link $L$. The state of the ensemble of small links has quantum numbers $|j_l,m_l,n_l\rangle$. Thus the total Hilbert space is $\otimes_{l}\oplus_{j_l}(\mathcal{H}_{j_l}\otimes\mathcal{H}_{j_l})=\oplus_{j_l}(\otimes_{l}\mathcal{H}_{j_l})\otimes(\otimes_{l}\mathcal{H}_{j_l})$.  Each $(\otimes_{l}\mathcal{H}_{j_l})$ factor can be decomposed into a sum of irreducible representations.
See Figure \ref{links}.
In this sum, each representation with spin $J$ (with $0\le J\le \sum_l j_l$) can appear several times.  We label with an index $\alpha$ the degenerate states in the $J$ representation. This defines the states
$|J,M,N,\alpha\rangle$, where the range of $\alpha$ depends on $J$ (as that of $M$ and $N$). Explicitly, 
\be
|J,M,N,\alpha\rangle=\sum_{\alpha_\pm}    i_{\alpha_+}^{m_l,M}i_{\alpha_-}^{n_l,N}   |j_l,m_l,n_l\rangle, \ee
where $\alpha={\alpha_+,\alpha_-}$ and $\alpha_\pm$ labels a basis $i_{\alpha_\pm}$ in the space of the intertwiners in the product of the representations of spin $j_l$ and $J$. Then the coarse grained link density matrix is 
\be
\langle J,M,N|\rho|J',M',N'\rangle =\sum_{\alpha,\alpha'}\ \langle J,M,N,\alpha|\psi\rangle \langle\psi|J',M',N',\alpha'\rangle. 
\ee

\subsubsection{Coarse graining graphs}

Any general coarse-graining is a combination of collecting nodes and
summing links (see Figure 4 for an example). 
We can therefore now bring together the two steps above and 
construct coarse grained graphs explicitly. 
\be
\rho(U_L,U'_L)=\sum_{\alpha,\beta} \int dU_l dU'_{l}\ \overline{\psi(U_l)}\psi(U'_l)\ 
D(U_l)^{j_l}_{m_l,n_l}D(U'_l)^{j'_l}_{m'_l,n'_l}
i_\alpha^{m_lm_L}
i_\beta^{n_ln_L}
i_\alpha^{m'_lm'_L}
i_\beta^{n'_ln'_L}
D(U_L)^{j_L}_{m_L,n_L}D(U'_L)^{j'_L}_{m'_L,n'_L}.
\ee

\begin{figure}
\centerline{\includegraphics[height=3.2cm]{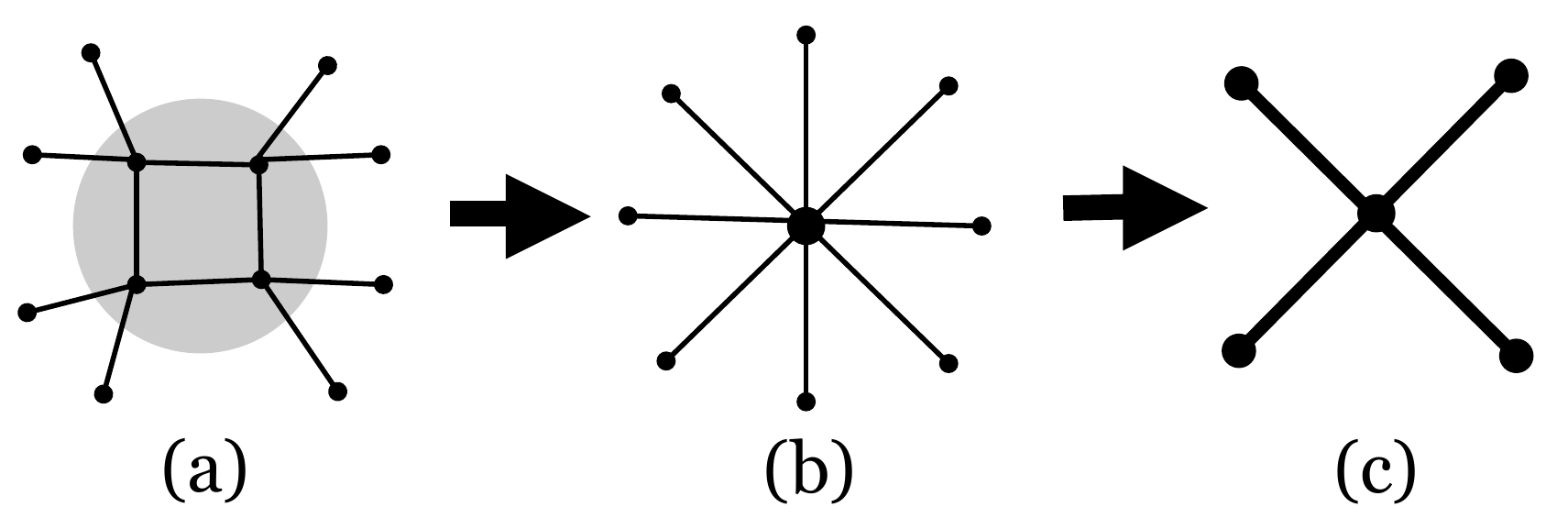}}
\caption{An example of a general coarse-graining which contains collecting-node and summing over links. Figure (a) is the fine graph we want to coarse-grained, starting by collecting all the nodes into single node in Figure (b). The next step is to sum the external links to obtain Figure (c). Note that summing links is done \textit{after} collecting nodes.}
\end{figure}

The steps above are the coarse graining from non-gauge invariant
Hilbert space to non-gauge invariant Hilbert space, namely, from $\mathcal{H}_{\gamma}\rightarrow\mathcal{H}_{\Gamma}$.
What we need is a coarse-graining from invariant subspace to invariant
subspace: $\mathcal{K}_{\gamma}\rightarrow\mathcal{K}_{\Gamma}$.
In order to do this, we need to give closure constraint (\ref{eq:3-3}) to each
Hilbert space. The easiest way is to project the density matrices
$\rho_{\gamma}\in\mathcal{H}_{\gamma}\otimes\mathcal{H}_{\gamma}^{*}$
and $\rho_{\Gamma}\in\mathcal{H}_{\Gamma}\otimes\mathcal{H}_{\Gamma}^{*}$ on each nodes $n$ and $N$ of the graph $\gamma$ and graph $\Gamma$, respectively: 
\begin{equation}
\rho_{\gamma}^{\left(\textrm{inv}\right)}=\pi_{n}\rho_{\gamma}\pi_{n},\label{eq:4-32}
\end{equation}
\begin{equation}
\rho_{\Gamma}^{\left(\textrm{inv}\right)}=\pi_{N}\rho_{\Gamma}\pi_{N},\label{eq:4-32-1}
\end{equation}.

The map from $\rho_{\gamma}^{\left(\textrm{inv}\right)}\in\mathcal{K}_{\gamma}\otimes\mathcal{K}_{\gamma}^{*}$
to $\rho_{\Gamma}^{\left(\textrm{inv}\right)}\in\mathcal{K}_{\Gamma}\otimes\mathcal{K}_{\Gamma}^{*}$ is the coarse-graining map.

\section{Correlation and entanglement entropy} \label{V}

The coarse-grained states, in general are mixed, thus, there is an
entropy related to the state. Given a mixed state 
\begin{equation}
\rho=\sum_{i}p_{i}\left|\psi_{i}\right\rangle \left\langle \psi_{i}\right|,\qquad\sum_{i}p_{i}=1,\label{eq:7-1}
\end{equation}
where $p_{i}$ is the probability (or the distribution) of state $\left|\psi_{i}\right\rangle ,$
the entropy of the state is defined as:
\begin{equation}
S=-\textrm{tr}\left(\rho\ln\rho\right),\label{eq:7-2}
\end{equation}
and bounded by:
\begin{equation}
S_{\textrm{min}}\leq S\leq S_{\textrm{max}}.\label{eq:7-3}
\end{equation}
The lower bound is $S_{\textrm{min}}=0,$ occurs when $\rho$
is pure, in other words, the distribution $p_{i}=\delta\left(i-i'\right)$,
is the Dirac-delta function which pick 1 specific microstate from all the possible
microstates. The upper bound is $S_{\textrm{max}}=\ln d$, occurs
when $\rho$ is a ``maximally-mixed'' state
\begin{equation}
\rho=\frac{1}{d}\left[\mathbb{I}\right],\label{eq:7-4}
\end{equation}
where $d$ is the dimension of the invariant subspace $\mathcal{K}_{\Gamma}$
and $\left[\mathbb{I}\right]$ is a $d\times d$ identity matrix.
All the possible reduced density matrix $\rho$ can be visualized
using a Bloch sphere, where the pure state lies in the surface of
the sphere, and the maximally-mixed state in the center of the sphere.
In between are the general mixed state. 

Notice that the entropy defined in this way does not measure the
classical ignorance of the fine details of the state of the system on 
the fine-grained graph $\gamma$.   Rather, it measures the quantum
correlation between the coarse-grained variables on the 
coarse-grained graph $\Gamma$ and the fine-grained variables.
In fact, is the full state was a product state, the resulting entropy 
would vanish. 

We can always view a mixed state as a state entangled with
a pure state of  larger system. In the case we are considering, the larger system is 
precisely the fine-grained graph. The coarse-grained state is obtained by tracing out 
information from the fine-grained graph. The entanglement
entropy measure how strong is the correlation \textit{within} the fine-grained 
graph, in other words, the correlation between the coarse graph and the ``details'' we
ignore. The correlation on spin-networks has been studied in \cite{Donnelly2012}.

\section{The geometry of the subset graph} \label{VI}

Let us now study the geometrical interpretation of the coarse grained states in ${\cal H}_{\Gamma}$.
These describe the geometry of connected polyhedra. The partition
that defines the subset graph $\Gamma$ is a coarse-graining of the
polyhedra into larger chunks of space. The surfaces that separate
these larger chunks of space are labelled by the big links $L$ and
are formed by joining the individual faces labelled by the links $l$
in $L$.

In general, it is clearly \emph{not} the case that the area $A_{L}$
is equal to the sum of the areas $A_{l}$ of all $l$ in $L$. However,
this \emph{is} the case if all these faces are parallel and have the
same orientation. Similarly, in general, it is clearly \emph{not}
the case that the volume $V_{N}$ is equal to the sum of the volumes
$V_{n}$ for the $n$ in $N$. However, this \emph{is} true if gluing
$n$ polyhedra one obtains a \emph{flat} polyhedron, with \emph{flat}
faces. This is because the formula $v=v(\vec{n}_{l})$ expresses the
evolve as a function of the boundary geometry of a region of space,
assuming flatness in the interior. Therefore the two operators
\begin{equation}
\triangle A_{L}:=\sum_{l\in L}A_{l}-A_{L}\label{eq:8-1}
\end{equation}
and 
\begin{equation}
\triangle V_{N}:=\sum_{n\in N}V_{n}-A_{N}\label{eq:8-2}
\end{equation}
provide a good measure of the failure of the geometry that the state
associates to $\Gamma$ to be flat, in the precise sense above. The
``other quantum numbers"   $\alpha$ mentioned in Section \ref{III} characterize whether the ``big grains of space" described by the states in ${\cal H}_{\Gamma}\otimes{\cal H}_{\Gamma}^{*}$
are actually ``flat" or not in this sense. Let the
coarse-graining given by the projection map as follow (supposing the trace over quantum numbers already included in the projection map):
\begin{equation}
\pi_{\gamma\Gamma}:\mathcal{K}_{\gamma}\otimes\mathcal{K}_{\gamma}^{*}\rightarrow\mathcal{K}_{\Gamma}\otimes\mathcal{K}_{\Gamma}^{*}.\label{eq:8-3}
\end{equation}
The map $\pi_{\Gamma\gamma}$ erases the geometrical information
at a scale smaller than the scale described by the coarse-grained
states on $\Gamma$, via the tracing over quantum number. Notice that in general, on the gauge invariant
states in $\mathcal{K}_{\gamma}\otimes\mathcal{K}_{\gamma}^{*}$
\begin{equation}
\pi_{n}\rho_{\psi}\pi_{n}=0,\quad\rho_{\psi}\in\mathcal{K}_{\gamma}\otimes\mathcal{K}_{\gamma}^{*},\label{eq:8-4}
\end{equation}
while gauge invariant states in $\mathcal{K}_{\Gamma}\otimes\mathcal{K}_{\Gamma}^{*}$
\begin{equation}
\pi_{N}\rho_{\Psi}\pi_{N}=0,\quad\rho_{\Psi}\in\mathcal{K}_{\Gamma}\otimes\mathcal{K}_{\Gamma}^{*},\label{eq:8-5}
\end{equation}
which implies that gauge invariance acts differently on the fine grained
and coarse-grained states. The states where $\pi_{N}\rho_{\Psi}\pi_{N}=0$
form a linear subspace of $\mathcal{H}_{\Gamma}\otimes\mathcal{H}_{\Gamma}^{*}$,
formed by states that are ``flat on each big node".

\begin{figure}
\centerline{\includegraphics[height=3.3cm]{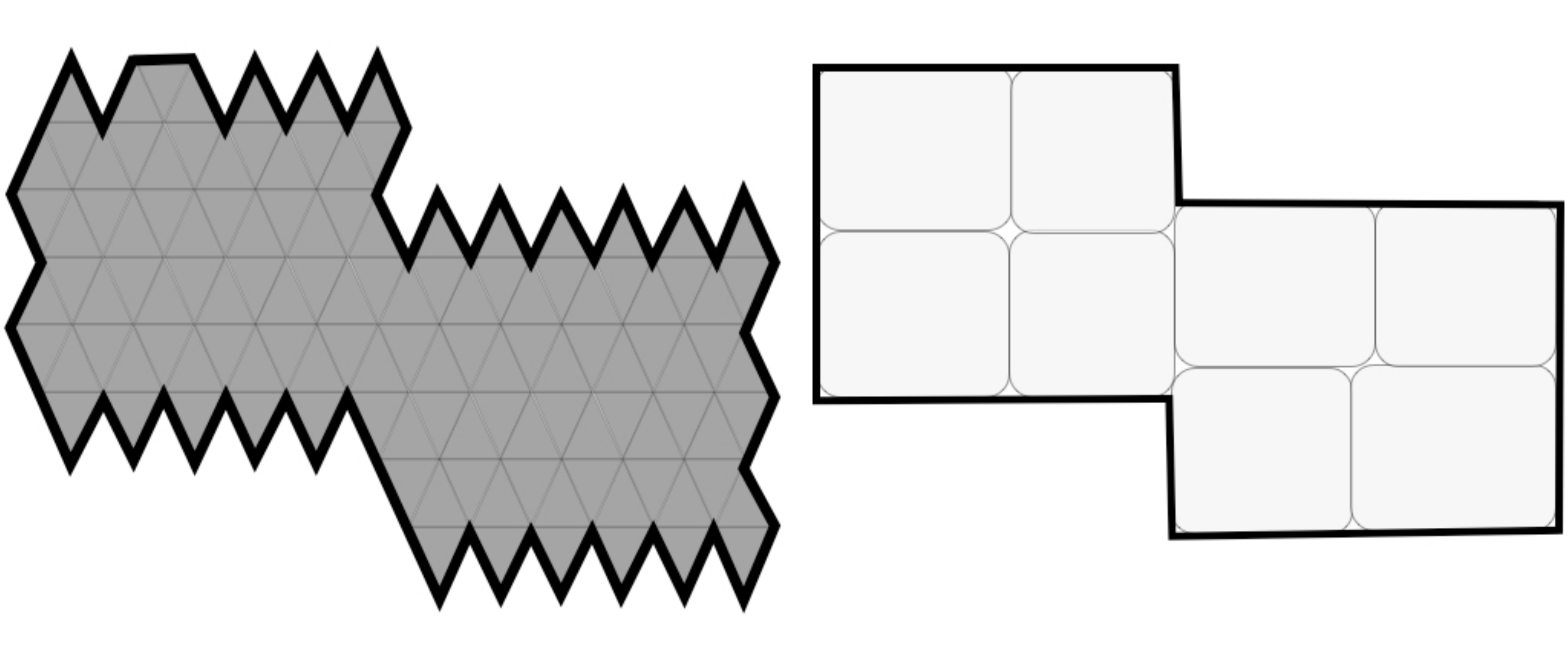}}
\caption{The coast of the fine grained discretization of the island is longer that the coast of the coarse-grained one (dual to a subset graph of the dual of the first discretisation). The differences measures the roughness of the coast. }
\end{figure}

A two-dimensional analog of this situation is illustrated in Figure
5, where we can associate two different lengths to the coast of the
black island: the fine grained length obtained by summing the small
black segments, while the coarse-grained length is the length of the
boundary of the coarse-grained discretization.

Finally, consider a family of graphs $\gamma_{m}$, with $m=0,...,M$,
such that each $\gamma_{m-1}$ is a subset graph of $\gamma_{m}$
and $\gamma_{M}=\gamma$. Call this a ``decomposition" of $\gamma$. The Hilbert spaces $\mathcal{H}_{\gamma_{m}}\otimes\mathcal{H}_{\gamma_{m}}^{*}$
are nested into one another, in the sense that there is a projection
\begin{equation}
\pi_{\gamma_{m}\gamma_{m-1}}:\mathcal{H}_{\gamma_{m}}\otimes\mathcal{H}_{\gamma_{m}}^{*}\rightarrow\mathcal{H}_{\gamma_{m-1}}\otimes\mathcal{H}_{\gamma_{m-1}}^{*}\label{eq:8-6}
\end{equation}
for each $m>0$. The set of area and volume operators $A_{L}^{m}$
and $V_{N}^{m}$ on each ${\cal H}_{\gamma_{m}}$ give a coarse grained
description of the geometry, which becomes finer as $m$ increases.
Each projection map erases ``other quantum numbers" which is related to the information of the curvature of spacetime. 

To have a good visualization of the coarse-grained geometries, it
is helpful to consider the classical picture. In the 4-dimensional
theory, the graph is defined at the boundary of a 3-dimensional hypersurface,
the spin operator on the links is related to the area operator by
$\vec{A}_{l}=8\pi\gamma G\left|\vec{J}_{l}\right|$. Given a 3-valent
graph with spins operators $\vec{J}_{l_{a}}$, $\vec{J}_{l_{b}}$,
and $\vec{J}_{l_{c}}$ on each links, the dihedral angle between $\vec{J}_{l_{b}}$
and $\vec{J}_{l_{c}}$ can be obtained from the angle operator, defined
by
\begin{equation}
\cos\hat{\theta}_{a}=\frac{\left|\vec{J}_{l_{b}}\right|^{2}+\left|\vec{J}_{l_{c}}\right|^{2}-\left|\vec{J}_{l_{a}}\right|^{2}}{2\left|\vec{J}_{l_{b}}\right|\left|\vec{J}_{l_{c}}\right|}.\label{eq:8-7}
\end{equation}
Applying this operator to the spin network state in the $\oplus$-basis
on each node, it gives the dihedral angle between $\vec{J}_{l_{b}}$
and $\vec{J}_{l_{c}}$ on the internal links $l_{b}$ and $l_{c}$: 
\begin{equation}
\theta_{a}=\cos^{-1}\left(\frac{j_{l_{b}}\left(j_{l_{b}}+1\right)+j_{l_{c}}\left(j_{l_{c}}+1\right)-j_{l_{a}}\left(j_{l_{a}}+1\right)}{2\sqrt{j_{l_{b}}\left(j_{l_{b}}+1\right)j_{l_{c}}\left(j_{l_{c}}+1\right)}}\right).\label{eq:deficit}
\end{equation}
The Regge intrinsic curvature of a discretized manifold is given by
the deficit angle on the hinges, the $(n-2)$-dimensional simplices
of the $n$-dimensional simplex. Thus, given a ``loop" graph with $n$-external
links, the deficit angle for a general $n$-polytope ($n$-valent
loop graph) is: 
\begin{equation}
\varepsilon=2\pi-\sum_{a}^{n}\theta_{a},\label{eq:regge2}
\end{equation}
where $\theta_{a}$ is the dihedral angle along the hinges. Using
the Regge curvature, we can study how area and volume, as quantum
observables, affected by spin network coarse-graining in the framework
of 4-dimensional theory.

\subsection{Coarse-grained area} \label{VIIIa}

Recall the boundary of spacetime, which is a 3-dimensional space.
Triangulation on the boundary is defined using flat polyhedra. Every
closed, flat, $n$-polyhedron satisfy the closure relation on the
node given by (\ref{eq:3-3}). Consider the net of a polyhedron
illustrated by Figure 6. 

\begin{figure}
\centerline{\includegraphics[height=5cm]{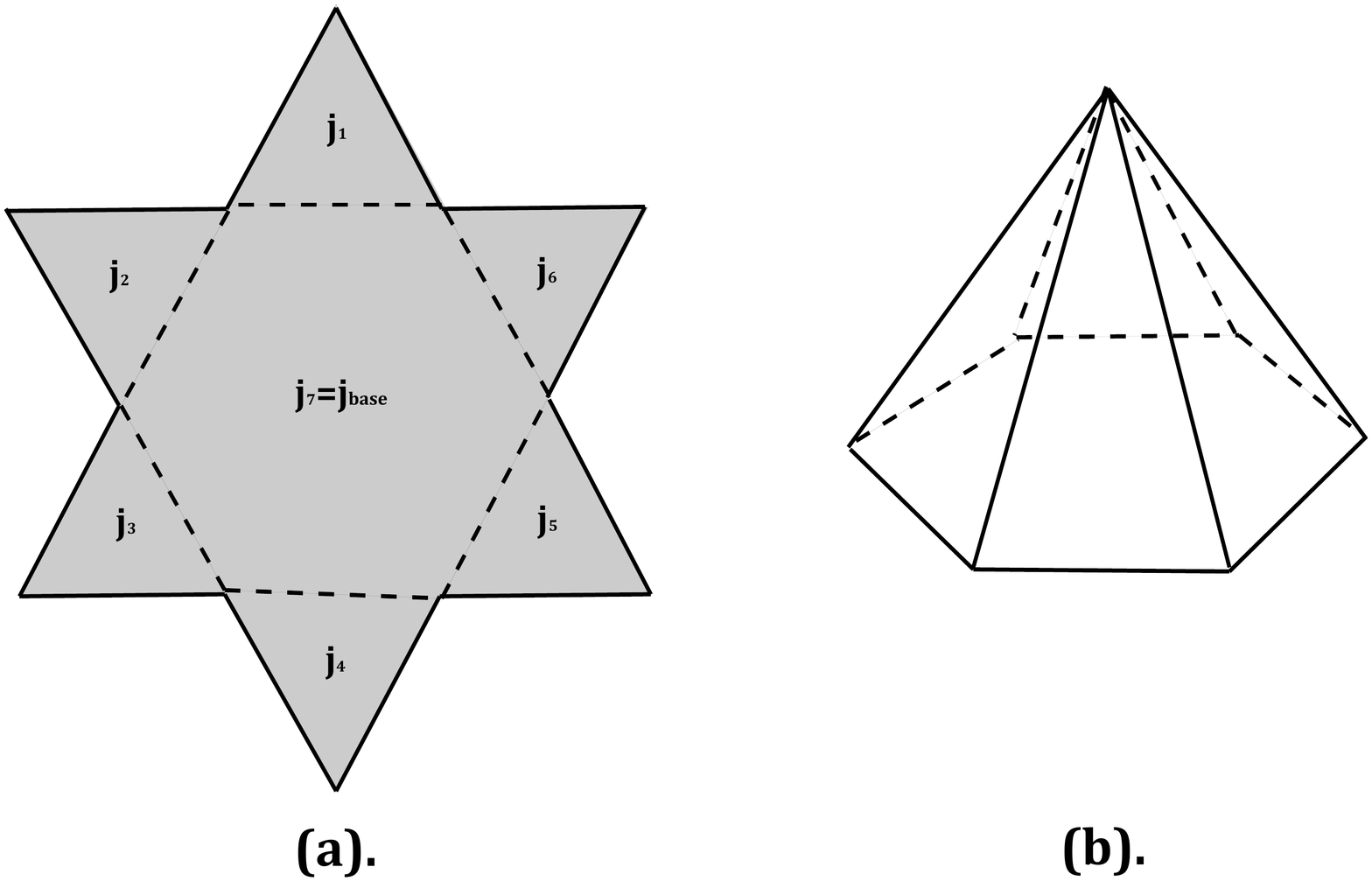}}
\caption{The net of a polyhedron. The ``hat'' is the triangles surrounding the
base, which is the polygon.}
\end{figure}

Suppose we only add the area operator on the ``hat'' (that is, to add
all the $\vec{J}_{l_{i}}$'s except the one at the ``base''). Since
the interior of polyhedron is flat, the closure relation can be written
as 
\begin{equation}
\vec{J}_{\textrm{base}}=-\sum_{i=1}^{n-1}\vec{J}_{l_{i}}.\label{eq:base}
\end{equation}
Then, the area operator on the base is 
\begin{equation}
\left|A_{\textrm{base}}\right|=8\pi\gamma G\left|\vec{J}_{\textrm{base}}\right|=8\pi\gamma G\left|-\sum_{i=1}^{n-1}\vec{J}_{l_{i}}\right|,\label{eq:abase}
\end{equation}
but this is the definition of the big link given by (\ref{eq:3-7}), and we
can defined the \textit{coarse-grained area} as: 
\begin{equation}
A_{L}=\left|A_{\textrm{base}}\right|=8\pi\gamma G\left|\vec{J}_{L}\right|\label{eq:coarsearea}
\end{equation}

Thus, for a 2-dimensional surface, we can always think the coarse
grained area $A_{L}$ as the area of the (flat) base of a polyhedron,
while the total sum of area $A_{l}$ is the area around the ``hat'',
i.e., the area of $n$ triangle which form the net of the polyhedron
in Figure 6.

We mentioned that the differences between the coarse-grained and the
fine-grained area gives a good measurement on how the space deviates
from being flat. Then, it is possible to obtain the explicit relation
between the Regge curvature with these area differences in some special
cases. To simplify, let the triangulation of the 2-dimensional surface
is defined in Figure 6, using $n$-\textit{isosceles} triangles (instead
of arbitrary triangles) with the angle between two isosceles lengths is $\theta$, satisfying $\theta+2\alpha=\pi$, and the length opposite to $\theta$ is $r$.
Then, the base is an $n$-polygon, and the triangulation of this portion
of surface is formed by $n$ triangles. The sum of the $n$-isosceles
triangle's area is the total surface area $\sum_{l}A_{l}$, 
\begin{equation}
\sum_{l}A_{l}=\frac{nr^{2}}{4}\cot\left(\frac{\theta}{2}\right),\label{eq:1}
\end{equation}
while the base ($n$-polygon) area is the coarse-grained area $A_{L}$
\begin{equation}
A_{L}=\frac{nr^{2}}{4}\cot\left(\frac{\pi}{n}\right).\label{eq:2}
\end{equation}

From (\ref{eq:1}) and (\ref{eq:2}) we obtain the dihedral angle
for one isosceles triangle 
\begin{equation}
\theta=2\cot^{-1}\left(\frac{\left|\sum_{l}A_{l}\right|}{\left|A_{L}\right|}\cot\left(\frac{\pi}{n}\right)\right).\label{eq:8-1}
\end{equation}
The Regge curvature for a 2-dimensional surface is defined as $2\pi$
minus the sum of all dihedral angle surrounding a point of the triangulation, which is $n\theta$. Finally, we obtain the Regge curvature
as a function of the coarse-grained and the fine-grained area 
\begin{equation}
\varepsilon=2\left(\pi-n\cot^{-1}\left(\frac{\left|\sum_{l}A_{l}\right|}{\left|A_{L}\right|}\cot\left(\frac{\pi}{n}\right)\right)\right).\label{eq:7}
\end{equation}

In the classical limit, it is clear that there can be states where
$\varepsilon=0$ or $\Delta A_{L}=0$. These correspond to geometries
where the normals to the facets forming the large surface $L$ are
parallel. However, this is only true in the classical limit, namely
disregarding Planck scale effects. If we take Planck-scale effects
into account, we have the remarkable result that 
\begin{equation}
\Delta A_{L}>0\label{eq:8-2}
\end{equation}
This can be seen as follows. Let $j_{l}$ be the spins associated
with the facets $l$. Then
\begin{equation}
\sum_{l}A_{l}=8\pi\gamma G\sum_{l}\sqrt{j_{l}(j_{l}+1)}.\label{eq:8-3}
\end{equation}
The total area of the large face is given by the Casimir of the operator
$\vec{J}_{L}$, which is the sum of the individual $j_{l}$ and lives
in the tensor product of the Hilbert spaces ${\cal H}_{\j_{l}}$.
Decomposing this tensor product into irreducible representations of
$SU(2)$, the highest possible representation appearing in
the product is the one with spin $J=\sum_{l}j_{l}$. Therefore the
maximum area of the large surface is 
\begin{equation}
A_{L}=8\pi\gamma G\hbar\sqrt{\sum_{l}j_{l}\left(\sum_{l}j_{l}+1\right)}.\label{eq:8-4}
\end{equation}
Thus 
\begin{eqnarray}
\Delta A_{L} & > & 8\pi\gamma G\hbar\sum_{l}\sqrt{j_{l}(j_{l}+1)}-8\pi\gamma G\hbar\sqrt{\sum_{l}j_{l}\left(\sum_{l}j_{l}+1\right)}\\
 & = & 8\pi\gamma G\hbar\sum_{l}\sqrt{j_{l}(j_{l}+1)}-8\pi\gamma G\hbar\sum_{l}\sqrt{j_{l}\left(j_{l}+\frac{j_{l}}{\sum_{l}j_{l}}\right)}>0.\nonumber 
\end{eqnarray}
unless there is a single non vanishing spin. Expanding for large spins
and keeping the first order, we have
\begin{equation}
\Delta A_{L}>4\pi\gamma\hbar G\, n\label{eq:8-5}
\end{equation}
where $n+1$ is the number of facets. 
Therefore \emph{the fine grained area is always strictly larger than the coarse grained area}. There is a Planck length square contribution
for each additional facet. It is as if there was an irreducible Planck-scale
fluctuation in the orientation of the facets.

\subsection{Coarse-grained volume} \label{VIIIb}

In the same manner as the surface's coarse-graining, we triangulate a 
3-dimensional chunk of space using $n$-symmetric tetrahedra. 
The Regge curvature is defined by the dihedral
angle on the \textit{bones} of the tetrahedra. Concretely, let's take
for instance a very symmetric situation, where all ``equatorial''
segments of the tetrahedron have the same length $E$, all the ``meridian'' segment length are 1, and the axis (the hinges bone) length is $h$.
For $n=3$ case, we have the illustration in Figure 7(a). 

\begin{figure}
\centerline{\includegraphics[height=4cm]{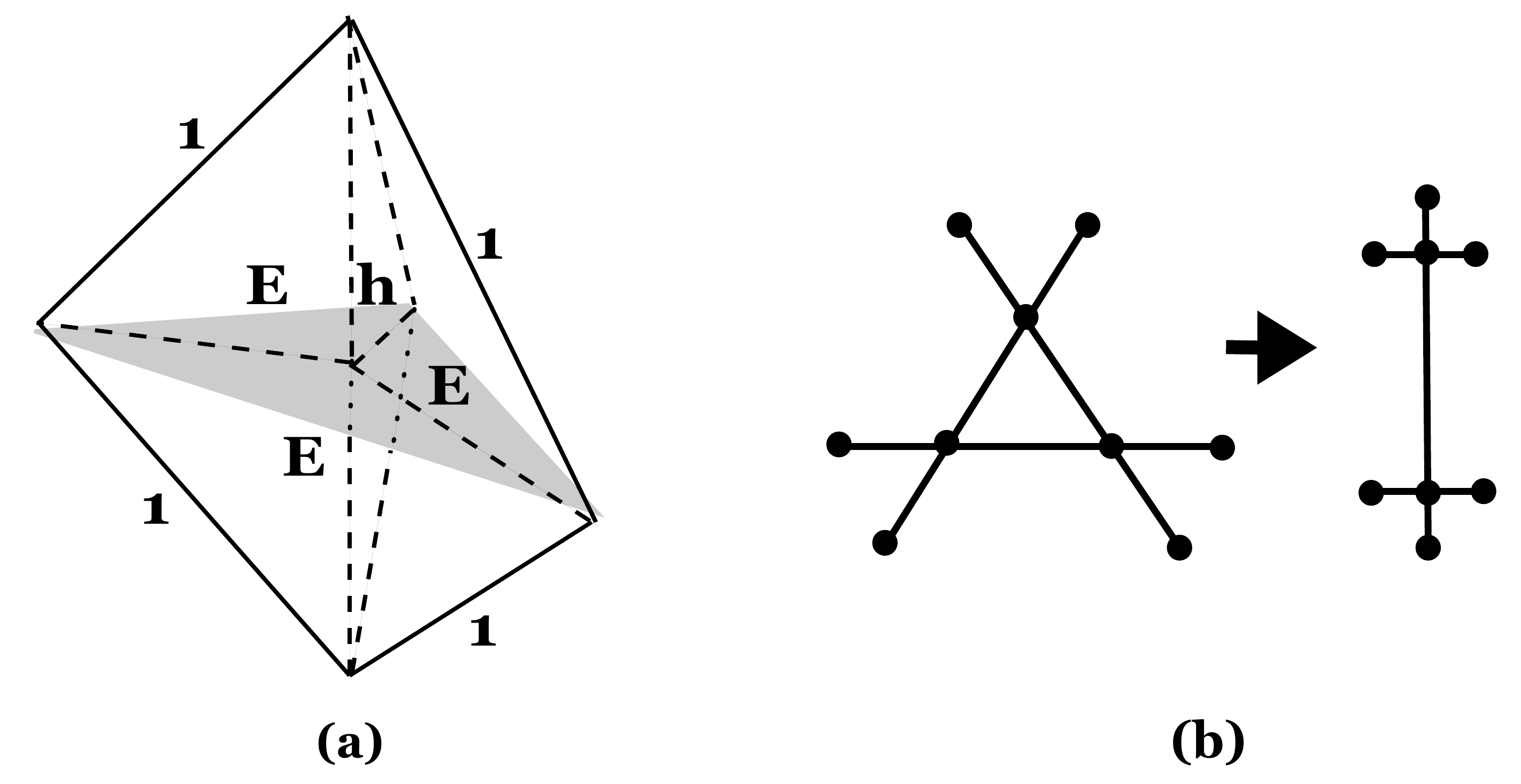}}
\caption{(a) A triangulation of a 3-dimensional chunk of space by 3 tetrahedra. (b) the inverse 2-3 Pachner move.}
\end{figure}

Using the volume of one tetrahedron, $V_{\textrm{tetra}}=\frac{Eh}{12}\sqrt{4-h^{2}}\cos\left(\frac{\theta}{2}\right)$,
we obtain the fine-grained volume, which is the volume of $n$ symmetric
tetrahedra: 
\begin{equation}
\sum_{n}V_{n}=\frac{nEh}{12}\sqrt{4-h^{2}}\cos\left(\frac{\theta}{2}\right).\label{eq:finevol}
\end{equation}
The coarse-grained volume is the volume of the 3-dimensional ``base'',
which is the volume of the $n$-``diamond'': 
\begin{equation}
V_{N}=\frac{nE^{2}h}{12}\cot\frac{\pi}{n}.\label{eq:coarsevol}
\end{equation}
The relation between the dihedral angle $\theta$ with $E$ and $h$
is 
\begin{equation}
\cot\left(\frac{\theta}{2}\right)=\frac{\sqrt{4-h^{2}}\cos\left(\frac{\theta}{2}\right)}{E},\label{eq:relation}
\end{equation}
and by combining (\ref{eq:finevol}), (\ref{eq:coarsevol}), and (\ref{eq:relation}),
we obtain $\cot\left(\frac{\theta}{2}\right)=\frac{\sum_{n}V_{n}}{V_{N}}\cot\left(\frac{\pi}{n}\right)$,
so, the Regge curvature is 
\begin{equation}
\varepsilon=2\left(\pi-n\cot^{-1}\left(\frac{\sum_{n}V_{n}}{V_{N}}\cot\left(\frac{\pi}{n}\right)\right)\right).\label{eq:regge3d}
\end{equation}

Notice that this is just a classical example. In quantum picture, adding two
quantum tetrahedra does not gives only a triangular bipyramid, it
could give other possible geometries which have 6 facets, i.e., a
\textit{parallelepiped}, or a \textit{pentagonal-pyramid}. 

The relation between the fine-grained volume $\sum_{n}V_{n}$ with
the coarse-grained volume $V_{N}$ is much more complicated. The coarse-grained
volume can be smaller \textit{or} bigger than the fine-grained volume, since there
are much more degrees of freedom than the 2-dimensional case concerning
the area. The algorithm to calculate the spectrum of the volume operator
acting on the state of a ``fuzzy'' tetrahedron is derived explicitly
at {[}Carlo{]}.

Consider a coarse-graining described by the inverse 2-3 Pachner move, see Figure 7(b). The fine-graph is related to 6-face polyhedron obtained by adding
three tetrahedra together. While the coarse-graph is related to a
6-facets obtained from only two tetrahedra. The spin network state
of the fine and coarse graph are given by $\left|j_{1},\ldots,j_{6},k_{a},k_{b},k_{c},l_{1},l_{2},l_{3}\right\rangle $
and $\left|j_{1},\ldots,j_{6},k_{a}',k_{b}',l\right\rangle $, respectively.
Given a fixed value for the quantum number on the external links $j_{1},\ldots,j_{6},$ we
still have degrees of freedom from the remaining quantum numbers (the
fine-graph has six quantum numbers on the internal links, while the
coarse-graph has three). These extra quantum numbers determine the
shape and the volume of the polyhedron.

For concreteness, let's fixed all the quantum number of the external
links to have value $j=\frac{1}{2}.$ Then, for the coarse-graph, the
remaining quantum number are $k_{a}',k_{b}'$ which determine the
volume of the two tetrahedra, say tetrahedron $a$ and $b$, and $l$,
which 'link' the two tetrahedra and determine the shape (or the volume,
these two things are related to each other) of the constructed polyhedron.
From the representation theory, we know that the possible value for
$k_{a}',k_{b}'$ is $\left\{ 0,1\right\} $ (because $\frac{1}{2}\otimes\frac{1}{2}=0\oplus1$),
and the possible value for $l$ is $\left\{ \frac{1}{2},\frac{3}{2}\right\} $
(because $\frac{1}{2}\otimes\frac{1}{2}\otimes\frac{1}{2}=\frac{1}{2}\oplus\frac{1}{2}\oplus\frac{3}{2}$).
Thus, there are six different combinations obtained from the spin
network state $\left|j_{1},\ldots,j_{6},k_{a}',k_{b}',l\right\rangle $
of the coarse-graph which are related to six possible 6-face polyhedron
with different volume. The volume spectrum can be obtained by calculating
the spectrum of each tetrahedra, then adding them together according
to the six possible combinations.

Similar with the coarse-graph, for the fine-graph, the remaining quantum
number are $k_{a},k_{b},k_{c}$ which determine the volume of the
three tetrahedra, say tetrahedra $a,b,$ and $c.$ And also $l_{1}$,
$l_{2}$ and $l_{3}$ which ``link'' tetrahedra $b$ and $c$, tetrahedra
$c$ and $a$, and tetrahedra $a$ and $c$, respectively. They also
determine the volume of the constructed polyhedron. But since there
are infinite possible combinations for $l_{1},l_{2},l_{3}$,
the polyhedron constructed by this ``bubble'' graph also varies infinitely,
without any upper bound to the volume.

According to the discussion above, it is clear that the coarse-grained
volume $V_{N}$ can be smaller or greater than the fine-grained volume
$\sum_{n}V_{n}$. There is no such irreducible fluctuations observed
in the 2-dimensional case concerning the area.

\subsection{Odd-face polyhedron} \label{VIIIc}

The building blocks for a 3-dimensional geometry is a ``fuzzy'' tetrahedron,
constructed from a 4-valent graph. Higher-valent graph related
to more complicated geometries can be constructed from these 4-valent
graphs by contracting  indices. But since a contraction
only concern two indices of the intertwiner at the same time, we could
only obtain even-valent intertwiners from a contraction of 4-valent
intertwiner, which means we could only have even-face polyhedron in
the theory. Then, how to obtain odd-face polyhedron? This problem
could be solved naturally by defining the procedure of coarse-graining
given in Section \ref{III}.

Let's consider a classical picture of a 5-face polyhedron. There are
two possible geometries related to this 5-face polyhedron: the ``pyramid''
and the ``truncated''-tetrahedron, and this had been studied in \cite{Haggard2013}.
In this classical case, we only consider the pyramid. Classically,
it is clear that we can always divide a pyramid into two tetrahedra.
But, conversely, to construct a pyramid from two arbitrary tetrahedra,
we need an additional constraint.

Consider a 6-valent graph arise from a contraction of two 4-valent
graph which are related to a 6-face polyhedron. Classically, the geometry
of this 6-valent graph is a \textit{triangular bipyramid}, a ``diamond'' (In a quantum picture, it does not need to be a pyramid. It could be
any possible geometries which have 6 faces, for example, a cube, etc.) A pyramid is a special case of a ``diamond'' where two of the face
are ``merged'' together to give one flat face with larger area. This
can be done by replacing two links on the dual space, by a ``bigger''
link, and the spin operator on the 'big' link is the sum of the two
operators on the ``smaller'' links, which we had already defined as
the addition of spin of the subset graph in (\ref{eq:3-7}). The total area related
to the ``big'' link is the \textit{coarse-grained area}. Thus, with
the definition of the subset graph and coarse-graining in Section
\ref{III}, we can obtain all arbitrary 3-dimensional geometries from
adding ``fuzzy'' tetrahedra.

\section{How many quanta of space are there in quantum spacetime?} \label{VII}

Armed with the observations of Section \ref{II} and with the technology
developed in Sections \ref{III} to \ref{VI}, let us return to the
question of the number of quanta in a given classical geometry.
The second example of Section \ref{II} (the oscillators) shows that
the number of quanta is not an absolute property of a quantum state:
it depends on the basis on which the state is expanded. In turn, this
depends on the way we are interacting with the system.
The first example (the photons) clarifies that the usual notion of
particle in quantum field theory refers to the quanta of the Fourier
modes. Each of these describe the aspect of the field that can interact
with an antenna of a given frequency, and captures non local and coarse
grained quantities of the electromagnetic field. For instance, an
antenna tuned into a wavelength $\lambda$ is insensitive to the high-frequency
components of the field, if any of these is excited.
We cannot treat the gravitational field in the same manner at all
scales, because Fourier analysis requires a background geometry, which
is, in general, not available in gravity. Sections \ref{III} to \ref{VI},
however, provide a viable alternative: areas and volumes of big links
and big nodes, $A_{L}$ and $V_{N}$, capture large scale features
of the field, and are insensitive to higher frequency components of
the field, in a way similar to the long wavelength Fourier modes.
In fact, notice that this is what we mean when we refer to macroscopic
areas and volume. The area of a table is not the sum of individual
areas of all microscopic elements of its boundary; it is the area
of a coarse-grained description of the table where the surface is
assumed to be flat, even if in reality it is not, at small scales.
When we measure the gravitational field, that is, geometrical quantities,
we routinely refer to its long wavelength modes. For instance, we
can measure the Earth-Moon distance with a laser. What we are measuring
is a non-local, integrated value of the gravitational field, in the
same manner in which an antenna measures a single wavelength of the
electromagnetic field.

{\em The quanta of the gravitational field we interact with, are those described
by the quantum numbers of coarse-grained operators like $A_{L}$ 
and $V_{N}$, not the maximally fine-grained ones.} 

Given a region, we can coarse-grain it to capture
large scale degrees of freedom. On the set of quantum states that
live on the graph $\gamma$, we can either measure the observables
$A_{l},V_{n}$ or the coarse-grained obervables $A_{L},V_{N}$. These
second correspond to ``lower frequency modes'' of the
field. Given a decomposition $\gamma_{m}$ of $\gamma$, the area
and volume operators $A_{L}^{m}$ and $V_{N}^{m}$ describe (with
redundancy) increasingly fine grained modes of the gravitational field.
Their quantum numbers are roughly analog of the number of photons on
a given Fourier mode.

Therefore we can begin to answer the question of the title. The number
of quanta we see in a system depends on the way we interact with it.
When interacting with a gravitational field at large scales we are
probing coarse grained features of space, which can be described by
the quantum numbers $J_{L},V_{N}$ of a coarse-grained graph $\gamma_{0}$.
Probing the field as shorter scales tests higher modes, which can
be described by more fine grained subset graphs $\gamma_{1},...,\gamma_{m}$.

The relevance of this construction for understanding the scaling the
dynamics, cosmology and black holes will be studied elsewhere.

\vspace{3em}

We thank Etera Livine, Hal Haggard, Mingyi Zhang and Tim Kittel for  discussions and advices. S.A. is supported by Directorate General of Higher Education of Indonesia postgraduate scholarship and the Bourse du Gouvernement Fran\c{c}ais No. 765844C.

\appendix
\section{Coupling $n$-spins}

Consider a coupling of $n$-spins with orthonormal basis $\left|j_{1},m_{1},\ldots,j_{n},m_{n}\right\rangle $
which span the Hilbert space $\mathcal{H}=\mathcal{H}^{j_{1}}\otimes\ldots\otimes\mathcal{H}^{j_{n}}.$
There is an isomorphism from the direct product representation of
the Hilbert space to its direct sum representation, given by:
\begin{equation}
\mathcal{H}^{j_{1}}\otimes\ldots\otimes\mathcal{H}^{j_{n}}=\mathcal{H}^{j_{\textrm{min}}}\oplus\ldots\mathcal{\oplus H}^{j_{\textrm{max}}}=\mathcal{H},\label{eq:a-1}
\end{equation}
with $j_{\textrm{max}}=\sum_{i=1}^{n}j_{1}$ and $j_{\textrm{min}}$
is the minimum value of combination of $\left\{ j_{1},\ldots j_{n}\right\} $
under substraction. The orthonormal basis of the direct sum representation
Hilbert space is obtain by transformation:
\begin{equation}
\left|j_{1\ldots n},m_{1\ldots n},j_{1},\ldots ,j_{n},j_{12},\ldots ,j_{1\ldots n-1}\right\rangle =\sum_{m_{1},\ldots,m_{n}}\imath_{j_{1\ldots n}m_{1\ldots n}j_{12}\ldots j_{1\ldots n-1}}^{m_{1}\ldots m_{n}}\left|j_{1},m_{1},\ldots,j_{n},m_{n}\right\rangle ,\label{eq:a-2}
\end{equation}
with
\begin{equation}
\imath_{j_{1\ldots n}m_{1\ldots n}j_{12}\ldots j_{1\ldots n-1}}^{m_{1}\ldots m_{n}}=\left\langle j_{1},m_{1},\ldots,j_{n},m_{n}\right.\left|j_{1\ldots n},m_{1\ldots n},j_{1},\ldots ,j_{n},j_{12},\ldots ,j_{1\ldots n-1}\right\rangle ,\label{eq:a-3}
\end{equation}
is the transformation coefficient, usually called as\textit{ intertwiner}.
$j_{1\ldots i}$ is the quantum numbers of total angular momentum
from coupling $i$-spins, i.e., $\left|j_{1}-j_{2}\right|\leq j_{12}\leq j_{1}+j_{2}$,
$\left|j_{12}+j_{3}\right|\leq j_{123}\leq j_{12}+j_{3}$, and so
on. The inverse transformation is:
\begin{equation}
\left|j_{1},m_{1},\ldots,j_{n},m_{n}\right\rangle =\sum_{\binom{j_{1\ldots n},m_{1\ldots n}}{j_{12},\ldots,j_{1\ldots n-1}}}\imath_{m_{1}\ldots m_{n}}^{j_{1\ldots n}m_{1\ldots n}j_{12}\ldots j_{1\ldots n-1}}\left|j_{1\ldots n},m_{1\ldots n},j_{1},\ldots ,j_{n},j_{12},\ldots ,j_{1\ldots n-1}\right\rangle ,\label{eq:a-4}
\end{equation}
with
\begin{equation}
\imath_{m_{1}\ldots m_{n}}^{j_{1\ldots n}m_{1\ldots n}j_{12}\ldots j_{1\ldots n-1}}=\left\langle j_{1\ldots n},m_{1\ldots n},j_{1},\ldots ,j_{n},j_{12},\ldots ,j_{1\ldots n-1}\right.\left|j_{1},m_{1},\ldots,j_{n},m_{n}\right\rangle .\label{eq:a-5}
\end{equation}

In the most general case, there is no restriction for $\left\{ j_{1},\ldots j_{n}\right\} ,$
they can have any value. In this case the direct sum representation
of the Hilbert space in general does not have zero representation:
\begin{equation}
\otimes^{l}\mathcal{H}^{j_{l}}=\mathcal{H}^{j_{\textrm{min}}}\oplus\ldots\mathcal{\oplus H}^{j_{\textrm{max}}},\quad j_{\textrm{min}}\geq0.\label{eq:a-6}
\end{equation}
A more special case occurs if we restrict the spins to satisfy Clebsch-Gordon
condition $j_{1\ldots n-1}\equiv j_{n}.$ This condition guarantees
$j_{\textrm{min}}=0,$ and the Hilbert space always has zero representation:
\begin{equation}
\otimes^{l}\mathcal{H}^{j_{l}}=\mathcal{H}^{0}\oplus\ldots\mathcal{\oplus H}^{j_{\textrm{max}}}.\label{eq:a-7}
\end{equation}
Moreover, we consider the invariant subspace of Hilbert space $\mathcal{H}$
which is the sum of all zero representations $\oplus^{d}\mathcal{H}^{0}.$
We called this subspace $\mathcal{K}=\oplus^{d}\mathcal{H}^{0}\subset\mathcal{H}$,
it has dimension $d$. The orthonormal basis in this space is obtained
by taking $j_{1\ldots n}=0$ in the $\oplus$-basis given by (\ref{eq:a-2}).
This cause $m_{1\ldots n}=0$. Together with condition $j_{1\ldots n-1}\equiv j_{n}$,
the orthonormal basis in $\mathcal{K}$ are: 
\begin{equation}
\left|j_{1\ldots n}=0,m_{1\ldots n},=0,j_{1},\ldots ,j_{n},j_{12},\ldots,j_{1\ldots n-2},j_{1\ldots n-1}=j_{n}\right\rangle =\left|0,0,j_{1},\ldots ,j_{n},j_{12},\ldots,j_{1\ldots n-2},j_{n}\right\rangle .\label{eq:a-8}
\end{equation}
The quantum number $j_{n}$ enters twice, so we can only write them
once, also we can omit the zeros, and the transformation (\ref{eq:a-2})
become
\begin{equation}
\left|j_{1},\ldots ,j_{n},j_{12},\ldots,j_{1\ldots n-2}\right\rangle =\sum_{m_{1},\ldots,m_{n}}\left\langle j_{1},m_{1},\ldots,j_{n},m_{n}\right.\left|j_{1},\ldots ,j_{n},j_{12},\ldots,j_{1\ldots n-2}\right\rangle \left|j_{1},m_{1},\ldots,j_{n},m_{n}\right\rangle ,\label{eq:a-9}
\end{equation}
or
\begin{equation}
\left|j_{1},\ldots ,j_{n},j_{12},\ldots,j_{1\ldots n-2}\right\rangle =\sum_{m_{1},\ldots,m_{n}}\imath_{00j_{12}\ldots j_{1\ldots n-2}}^{m_{1}\ldots m_{n}}\left|j_{1},m_{1},\ldots,j_{n},m_{n}\right\rangle ,\label{eq:a-10}
\end{equation}
using $\imath_{00,j_{12}\ldots j_{1\ldots n-2}}^{m_{1}\ldots m_{n}}$
as the invariant intertwiner, i.e., transformation coefficient which
transform the basis in the invariant subspace. The dimension of $\mathcal{K}$
(which is $d)$ depends on the degeneracies of the quantum number
$\left\{ j_{12},\ldots,j_{1\ldots n-2}\right\} .$ Fixing the spins
$\left\{ j_{1},\ldots j_{n}\right\} ,$ we can write general state
living in $\mathcal{K}$:
\begin{equation}
\left|\psi_{\textrm{inv}}\right\rangle =\sum_{j_{12},\ldots,j_{1\ldots n-2}}C_{j_{12},\ldots,j_{1\ldots n-2}}\left|j_{1},\ldots ,j_{n},j_{12},\ldots,j_{1\ldots n-2}\right\rangle \in\mathcal{K}.\label{eq:a-11}
\end{equation}

\section{Spin network basis}

Given a graph at the boundary of a discretized manifold, we associate
each link $l=\left(n,n'\right)$ of the graph by a group variable $U_{nn'}\in SU(2)$
and algebra variable $J_{nn'}\in\mathfrak{su(2)}^{*}$. Together, they form
an element of phase space on the boundary, $\left(U_{nn'},J_{nn'}\right)\in T^{*}SU(2).$
We labeled the link by spin-$j$, the representation of $SU(2)$ in
$(2j+1)$-dimension. The representation space is the Hilbert space of
the spin network, build over $SU(2),$ namely $\mathcal{H}=L^{2}\left[SU\left(2\right)\right]$,
one per each link. Since $SU(2)$ is a matrix group, it is an endomorphism
on $\mathcal{H}_{j}$, so it can by written by $\mathcal{H}=\mathcal{H}_{j}\otimes\mathcal{H}_{j}^{*}$.
Thus, the basis in this Hilbert space is
\begin{equation}
\left|j,n\right\rangle \left\langle j,n'\right|\in\mathcal{H}_{j}\otimes\mathcal{H}_{j}^{*},\label{eq:B1}
\end{equation}
each spin basis related to one end of the link (see Figure 8a).

\begin{figure}
\centerline{\includegraphics[height=3.5cm]{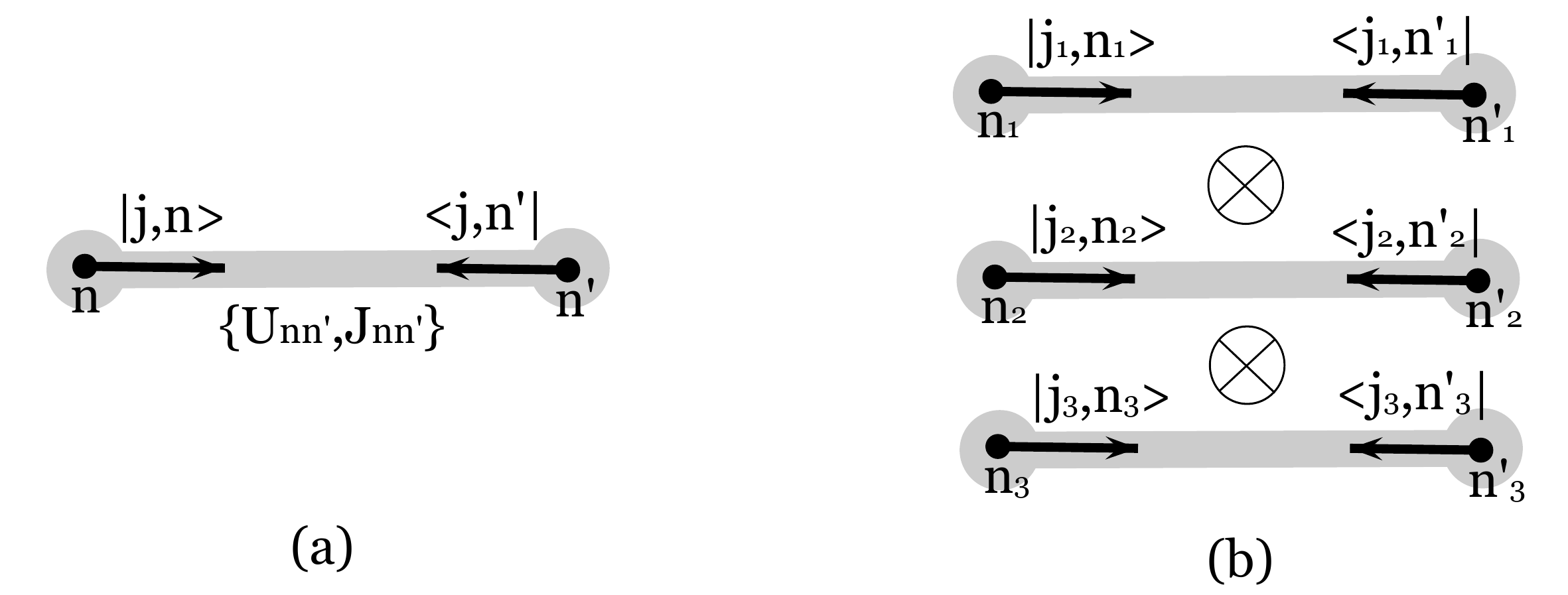}}
\caption{(a) Variables and spin quantum numbers attached to one link of a graph. (b) A system consisting three non-connected links. }
\end{figure}

The basis
$\left|j,n\right\rangle \left\langle j,n'\right|$ usually written
as $\left|j,n,n'\right\rangle ,$ and throughout this paper we will
use this notation. The general vector state living in the Hilbert
space $\mathcal{H}$ is
\begin{equation}
\left|\psi\right\rangle =\sum_{j,n,n'}\left\langle j,n,n'|\psi\right\rangle \left|j,n,n'\right\rangle ,\label{eq:B2}
\end{equation}
and the wave function $\psi\left(U_{nn'}\right)$ can be obtained
by contracting the vector state with the basis in the group representation:
\begin{equation}
\psi\left(U_{nn'}\right)=\left\langle U_{nn'}\mid\psi\right\rangle =\sum_{j,n,n'}\left\langle j,n,n'|\psi\right\rangle D_{nn'}^{j}\left(U_{nn'}\right),\label{eq:B3}
\end{equation}
with $D_{nn'}^{j}\left(U_{nn'}\right)=\left\langle U_{nn'}\mid j,n,n'\right\rangle $,
is the component of the Wigner-D matrix, or the \textit{rotator}.

\begin{figure}
\centerline{\includegraphics[height=4.2cm]{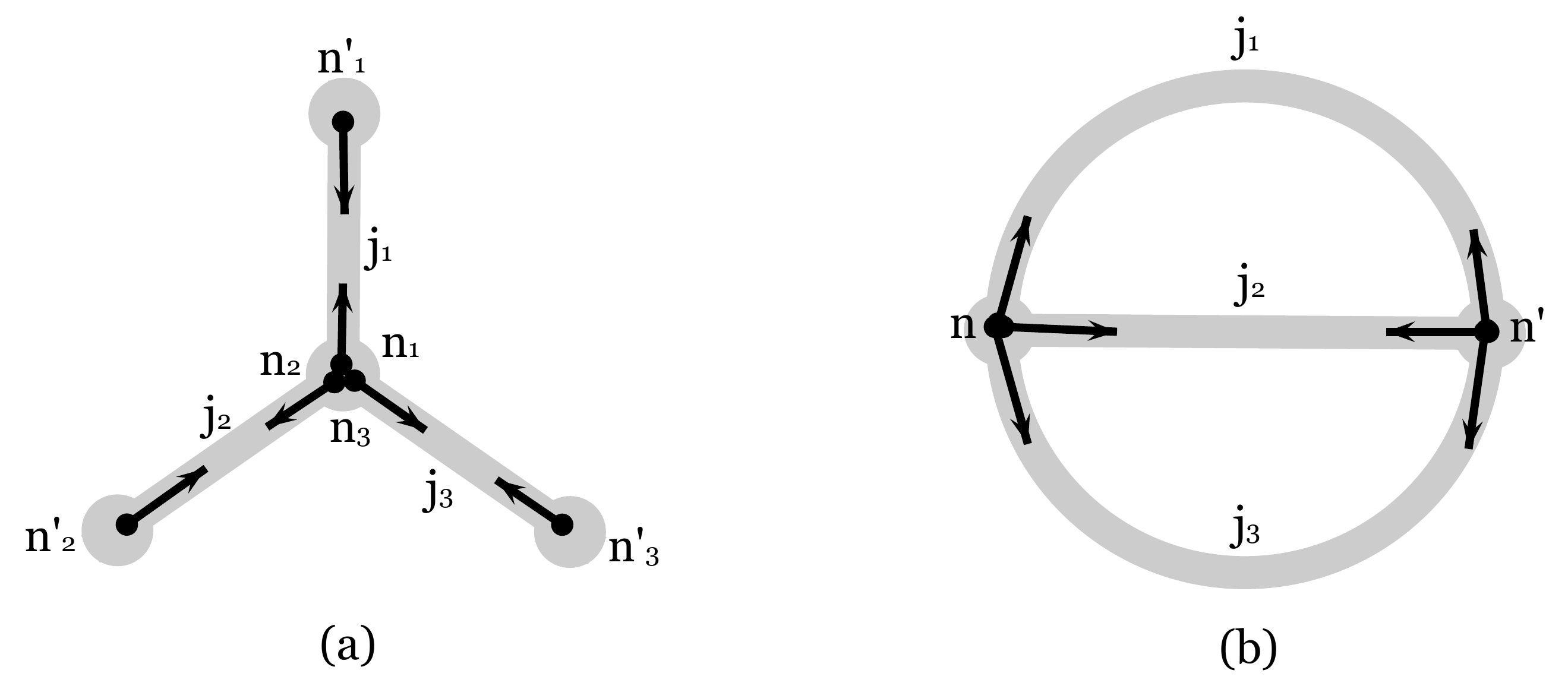}}
\caption{(a) A 3-valent graph. (b) A \textit{theta} graph.}
\end{figure}

Given a system with three links (see Figure 8b), we can write down its basis
as a tensor product of (\ref{eq:B1}):
\begin{equation}
\left|j_{1},n_{1},n'_{1}\right\rangle \otimes\left|j_{2},n_{2},n'_{2}\right\rangle \otimes\left|j_{3},n_{3},n'_{3}\right\rangle =\left|j_{1},n_{1},n'_{1}\right\rangle \left|j_{2},n_{2},n'_{2}\right\rangle \left|j_{3},n_{3},n'_{3}\right\rangle ,\label{eq:B4}
\end{equation}
we called this basis as the '$\otimes$-\textit{basis}'. Now, we arrange these
three links to form a 3-valent graph in Figure 9a. At node $n$, where the
links met, it is exactly a coupling of $\left|j_{1},n_{1}\right\rangle $,
$\left|j_{2},n_{2}\right\rangle $, and $\left|j_{3},n_{3}\right\rangle .$
So the basis of this $3$-valent graph is:
\begin{equation}
\left|j_{123},m_{123},j_{12},j_{1},j_{2},j_{3},n'_{1},n'_{2},n'_{3}\right\rangle =\sum_{n_{1},n_{2},n_{3}}i_{j_{123}m_{123}j_{12}}^{n_{1}n_{2}n_{3}}\left|j_{1},n_{1},n'_{1}\right\rangle \left|j_{2},n_{2},n'_{2}\right\rangle \left|j_{3},n_{3},n'_{3}\right\rangle ,\label{eq:B5}
\end{equation}
 where we couple $\left|j_{1},n_{1}\right\rangle $, $\left|j_{2},n_{2}\right\rangle $,
and $\left|j_{3},n_{3}\right\rangle $ using the intertwiner $i$
on the node, defined in Appendix A. We called basis in (\ref{eq:B5})
the '$\oplus$-\textit{basis}'. Clearly, (\ref{eq:B5}) is only a transformation
from $\otimes$-basis to $\oplus$-basis at node $n$.

For the spin network state, we require the $SU(2)$ gauge invariance on each node (because the graph is dual to flat quanta of space),
which means $\left|j_{1},n_{1}\right\rangle $, $\left|j_{2},n_{2}\right\rangle $,
and $\left|j_{3},n_{3}\right\rangle $ must satisfy Clebsch-Gordon
condition $j_{12}=j_{3}.$ This guarantees the $\oplus$-basis has
zero spin representation. The gauge invariant states are the state living
in this zero representation space, which are the state satisfying:
\begin{equation}
j_{123}=0,\; m_{123}=0.\label{eq:B6}
\end{equation}
Thus, the basis in the invariant subspace $\mathcal{K}\subset\mathcal{H}$
is:
\begin{equation}
\left|0,0,j_{3},j_{1},j_{2},j_{3}\right\rangle =\left|j_{1},j_{2},j_{3}\right\rangle =\sum_{n_{1},n_{2},n_{3}}i_{00j_{3}}^{n_{1}n_{2}n_{3}}\left|j_{1},n_{1},n'_{1}\right\rangle \left|j_{2},n_{2},n'_{2}\right\rangle \left|j_{3},n_{3},n'_{3}\right\rangle ,\label{eq:B7}
\end{equation}
by omitting the $j_{3}$ which appears twice, and not considering
the $n'$'s nodes on the left hand side. This procedure can be easily
generalized to any higher-valent graph. For example, is the 4-valent
graph, where the $\oplus$-basis of the invariant subspace is:
\begin{equation}
\left|j_{1},j_{2},j_{3},j_{4},\alpha\right\rangle =\sum i_{00\alpha j_{4}}^{n'_{1}n'_{2}n'_{3}n'_{4}}\left|j_{1},n_{1},n'_{1}\right\rangle \left|j_{2},n_{2},n'_{2}\right\rangle \left|j_{3},n_{3},n'_{3}\right\rangle \left|j_{4},n_{4},n'_{4}\right\rangle ,\label{eq:B8}
\end{equation}
with $j_{123}=j_{4},$ $j_{1234}=0$, $m_{1234}=0$, and $j_{12}=\alpha$.
The geometrical interpretation of this basis in 4-dimensional quantum
gravity, is the basis state of quantum tetrahedron, which also can
be written in the spin network basis:
\begin{equation}
\left|j_{1},j_{2},j_{3},j_{4},v_{n}\right\rangle ,\label{eq:B9}
\end{equation}
with each spins $j_{i}$ on the links are the quantum number of area
of four triangles forming the tetrahedron, and $v_{n}$ is the quantum
number of the volume. It is related to the quantum number $\alpha$
in the $\oplus$-basis.

As an example, we consider the \textit{theta} graph in Figure 9b, its $\oplus$-basis,
transforming from the $\otimes$-basis, is:
\begin{equation}
\left|j_{123},n_{123},n'_{123},\alpha,\beta,j_{1},j_{2},j_{3}\right\rangle =\sum_{\binom{n{}_{1},n{}_{2},n{}_{3}}{n'_{1},n'_{2},n'_{3}}}i_{j_{123}n_{123}\alpha}^{n{}_{1}n{}_{2}n{}_{3}}i_{j_{123}n_{123}\beta}^{n'_{1}n'_{2}n'_{3}}\left|j_{1},n_{1},n'_{1}\right\rangle \left|j_{2},n_{2},n'_{2}\right\rangle \left|j_{3},n_{3},n'_{3}\right\rangle ,\label{eq:B10}
\end{equation}
where $j_{123}=j_{123}^{\left(n\right)}=j_{123}^{\left(n'\right)}$,
$j_{12}^{\left(n\right)}=\alpha$, and $j_{12}^{\left(n'\right)}=\beta$.
The gauge invariant basis can be obtain by taking $j_{123}=0$, $n_{123}=n'_{123}=0$
and $\alpha=\beta=j_{3}$.

To conclude, the spin-network basis gives the information about which
spins that could be added together, and which could not. Spins pointing
out from the same node can be added together (represented by the $\oplus$-basis,
which is the total spin basis). We can freely transform $\otimes$-basis
to $\oplus$-basis and vice versa only on spins attached at the same
node, in the full, non-gauge invariant Hilbert space $\mathcal{H}$.


   
   \end{document}